\documentclass[sigconf,natbib=false]{acmart}
\AtBeginDocument{%
  }

\setcopyright{acmlicensed}
\copyrightyear{2025}
\acmYear{2025}
\acmDOI{XXXXXXX.XX?XXXX}
\acmConference[MobiHoc'25]{International Symposium on Theory, Algorithmic Foundations, and Protocol Design for Mobile Networks and Mobile Computing}{Oct 27--30,
  2025}{Houston, TX}
\acmISBN{978-1-4503-XX?X-X/2025/10}

\RequirePackage[
  datamodel=acmdatamodel,
  style=acmnumeric,
  ]{biblatex}

\begin{document}

\title{Demonstrating Visual Information Manipulation Attacks in Augmented Reality: A Hands-On Miniature City-Based Setup}

\author{Yanming Xiu}
\email{yanming.xiu@duke.edu}
\orcid{0009-0008-1547-3261}
\affiliation{%
  \institution{Duke University}
  \city{Durham}
  \state{NC}
  \country{USA}
}

\author{Maria Gorlatova}
\email{maria.gorlatova@duke.edu}
\affiliation{%
  \institution{Duke University}
  \city{Durham}
  \state{NC}
  \country{USA}
  }

\renewcommand{\shortauthors}{Xiu et al.}

\begin{abstract}
Augmented reality (AR) enhances user interaction with the real world but also presents vulnerabilities, particularly through Visual Information Manipulation (VIM) attacks. These attacks alter important real-world visual cues, leading to user confusion and misdirected actions. In this demo, we present a hands-on experience using a miniature city setup, where users interact with manipulated AR content via the Meta Quest 3. The demo highlights the impact of VIM attacks on user decision-making and underscores the need for effective security measures in AR systems. Future work includes a user study and cross-platform testing.
\end{abstract}





\maketitle

\section{Introduction}

Augmented reality (AR) is becoming increasingly popular, with a wide range of applications helping users interact with the real world in innovative ways. However, despite its advantages, AR also presents vulnerabilities when applied in real-world scenarios~\cite{VIM01, VIM03}. One such threat is the visual information manipulation (VIM) attack~\cite{vimsense, chen2025neurosymbolic}, which involves altering real-world visual elements in the AR environment in a detrimental way. By manipulating key visual cues, such as labels on buildings, road signs, or warning instructions, these attacks can mislead users, leading to incorrect decisions or unsafe behavior. The consequences of VIM attacks can be particularly severe in safety-critical domains such as autonomous driving and healthcare, where precise spatial awareness is crucial.

In our recent work, we discussed the taxonomy of VIM attacks and introduced VIM-Sense~\cite{vimsense}, a system that utilizes vision-language models (VLMs) and optical character recognition (OCR) modules to understand the semantic information in AR scenes and detect VIM attacks. While the concept and detection method of VIM attacks have been established, their impact on user behavior has not been fully explored. This demo provides a hands-on experience of VIM attacks, enabling users to interact with manipulated real-world elements in a controlled AR environment. Through the use of a miniature city setup, where real-world elements such as road signs and building labels are manipulated, users can directly experience the confusion and disorientation caused by VIM attacks. The demo provides an opportunity to analyze the impact of these attacks, showcasing how such manipulations can influence users' understanding of their environment. The primary goal of this demo is to raise awareness about the vulnerabilities of AR systems to VIM attacks and to emphasize the importance of developing effective detection and mitigation mechanisms that can preserve the integrity and reliability of AR experiences.

\section{Demonstration Setup}

\subsection{Miniature City Setup}

The core of the demo is a miniature city constructed on a corkboard, which is 90 cm in length and 60 cm in width. The city features a variety of real-world objects, such as toy buildings, trees, landmarks, and a road network, which together create an interactive and realistic small-scale urban environment. These elements are strategically placed to form a navigable cityscape. At the center of the setup is a mini toy car, equipped with remote control functionality, which allows users to drive it through the city. We selected a model that offers precise control and can perform both straight and turning maneuvers in narrow spaces. The user's task is to navigate the car to specific locations within the city, such as the "hospital," by following the road and interacting with real-world landmarks. The users are also reminded to adhere to real-life traffic rules, such as stopping at stop signs and following directional signs on the road.

\begin{figure*}[t]
\includegraphics[width=0.95\linewidth]{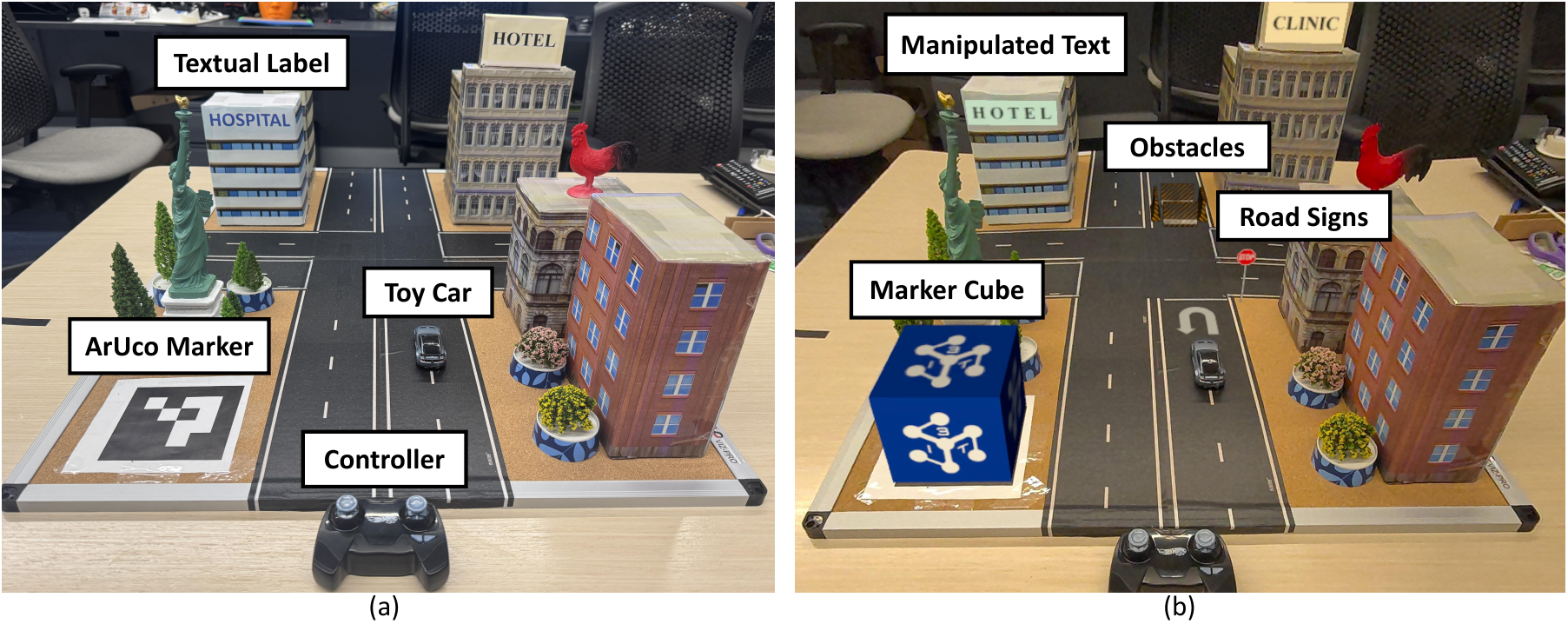}
\centering
\vspace{-0.4cm}
\caption{Overview of the miniature city demo setup. (a) shows the initial scene with real-world elements; (b) illustrates the user's view of the AR-manipulated scene, with several VIM attacks.}
\label{fig:vis}
\vspace{-0.5cm}
\end{figure*}

\subsection{VIM Attack Integration}

In this demo, we chose the Meta Quest 3 as the AR device due to its camera access~\cite{MetaQuest3Passthrough2025} and robust tracking capabilities, which are crucial for overlaying virtual content onto the real-world objects in the mini-city setup. The Quest 3's passthrough camera allows for a combination of virtual content and real-world elements. Additionally, its ability to integrate with ArUco marker tracking~\cite{metaaruco} ensures precise alignment of the virtual content with the real-world objects, making it an ideal choice for demonstrating VIM attacks. The attacks are designed to manipulate key visual elements that users rely on to navigate the environment:

\noindent \textbf{Manipulated Text}: Real-world buildings in the miniature city, such as the "hospital," have their text on the labels manipulated. For instance, the "hospital" sign is replaced with "hotel", while "hotel" is changed to "clinic." These alterations mislead the user and can cause confusion, as they may lead the user to the wrong locations.

\noindent \textbf{Road Signs}: Virtual road signs, such as U-turn signs on the road and stop signs at the intersection, are placed into the road network. These signs misdirect the user into taking incorrect turns or making inappropriate stops, leading them to incorrect destinations or causing them to miss their intended path.

\noindent \textbf{Obstacles}: Virtual road obstacles, such as concrete roadblocks, are inserted into the driving path, inducing the user to make decisions about how to navigate around them. These obstacles further simulate the disruptions caused by VIM attacks, forcing users to reassess their actions based on misleading or obstructed visual cues.

Through these manipulations, the demo highlights how VIM attacks can subtly distort the user’s perception of the real world, leading to confusion, misdirection, and incorrect decisions, making this demo a tool for demonstrating the potential dangers of VIM attacks in AR systems.

\vspace{-0.2cm}

\section{Future Work}

The demonstration provides an initial hands-on experience of VIM attacks, but there are several areas that require further exploration. Firstly, we will conduct a user study to quantitatively assess how VIM attacks influence user decision-making and behavior in AR environments. By analyzing how users interact with manipulated visual cues, we can gain insights into how these attacks disrupt their ability to navigate and complete tasks. In a pilot study, we tested the setup and app with three users and two of them did not realize the text on the building was manipulated, and drove to the hospital when asked to drive to the hotel. Another future direction is cross-platform testing: while this demo uses the Meta Quest 3, it is important to investigate how VIM attacks manifest on other AR devices, such as the Apple Vision Pro, which is generally regarded for its superior rendering capabilities. By addressing these areas, we aim to refine our understanding of VIM attacks and contribute to the development of more robust AR systems that can mitigate these vulnerabilities in real-world AR applications.

\vspace{-0.2cm}

\subsection*{Acknowledgements}
This work was supported in part by NSF grants CSR-2312760, CNS-2112562, and IIS-2231975, NSF CAREER Award IIS-2046072, NSF NAIAD Award 2332744, a Cisco Research Award, a Meta Research Award, Defense Advanced Research Projects Agency Young Faculty Award HR0011-24-1-0001, and the Army Research Laboratory under Cooperative Agreement Number W911NF-23-2-0224. The views and conclusions contained in this document are those of the authors and should not be interpreted as representing the official policies, either expressed or implied, of the Defense Advanced Research Projects Agency, the Army Research Laboratory, or the U.S. Government. This paper has been approved for public release; distribution is unlimited. No official endorsement should be inferred. The U.S.~Government is authorized to reproduce and distribute reprints for Government purposes notwithstanding any copyright notation herein.

\vspace{-0.1cm}

\printbibliography




\end{document}